\documentclass{PoS}

\title{Accidental permutation symmetries as a test for Grand Unification: the supersymmetric $SU(5)$ case}
\ShortTitle{Accidental permutation symmetries as a test for Grand Unification: SUSY $SU(5)$}

\author{Sylvain Fichet\\
        ICTP South American Institute for Fundamental Research \& Instituto de F\'isica Te\'orica, Universidade Estadual Paulista, S\~ao Paulo, Brazil }

\author{\speaker{Bj\"orn Herrmann}\\
LAPTh, Universit\'e Savoie Mont Blanc, CNRS, 9 Chemin de Bellevue, F-74941 Annecy-le-Vieux, France}

\abstract{
Unification of matter fields implies the existence of accidental permutation symmetries, which potentially remain immune to large quantum corrections up to the TeV scale. We investigate the case of a supersymmetric $SU(5)$ grand unified theory, where such a permutation symmetry is present in the up-type squark sector. We present a variety of tests allowing to challenge the $SU(5)$ hypothesis based on the observation of squarks at the LHC. These tests appear as relations among observables involving flavour-violating or chirality-flipping decays of squarks. Moreover, they rely on top-polarimetry and charm-tagging. As an example, we discuss the application to the scenario of Natural Supersymmetry, while more examples can be found in the related journal publications. }

\FullConference{38th International Conference on High Energy Physics\\
		3-10 August 2016\\
		Chicago, USA}

\begin{document}

% =================================================
\section{Introduction}

A major motivation to believe that electroweak and strong interactions are unified  at some high-energy scale is the fact that the Standard Model matter fields fit into complete representations of the $SU(5)$ gauge group  \cite{Georgi1974}, which may directly be broken into the Standard Model gauge group $SU(3) \times SU(2) \times U(1)$ \cite{SU5Refs}. Unraveling whether or not Nature is unified at some higher scale, possibly in a symmetry group containing $SU(5)$,  is a challenging open problem of particle physics. Whenever matter fields are unified into representations of the unified group, accidental permutation symmetries are expected to be present in the matter sector, at the unification scale. The meaning of ``accidental'' is that the permutation symmetry is present at the level of the renormalizable Lagrangian, but can possibly be violated by higher-dimensional operators.

We focus on the case of supersymmetric $SU(5)$ Grand Unified Theory (GUT). In a supersymmetric theory, the GUT representations include the superfields accounting also for the superpartners of the Standard Model degrees of freedom. A first consequence are two relations between the Yukawa and trilinear couplings of the down-type (s)quarks and the (s)leptons, $y_d^t ~=~ y_{\ell}$ and $a_d^t ~=~ a_{\ell}$, respectively. Note that these relations are exact at the $SU(5)$ unification scale. However, they are heavily spoiled when evolving the theory to lower energy scales through renormalization group running, such that they do not turn out to be useful to test the $SU(5)$ hypothesis based on, e.g., LHC observables.

In addition, a second set of relations holds within the sector of up-type (s)quarks,
\begin{equation}
	y_u^t ~=~ y_u \, \quad {\rm and} \quad a_u^t ~=~ a_u \,,
	\label{Eq:SU5uGUT}
\end{equation}
again valid exactly at the $SU(5)$ unification scale. These are the permutation symmetries of our focus. Interestingly, these relations remain almost stable to quantum corrections when evolving the parameters towards the TeV scale. Since the beta functions of $y_u$ and $a_u$ are dominated by symmetric terms, the introduced asymmetry remains small such that we have, e.g.,
\begin{equation}
	a_u^t ~\approx~ a_u \,.
	\label{Eq:SU5uTeV}
\end{equation}
at the TeV scale.  We have numerically shown that the introduced asymmetry is typically of the order of a few percent \cite{JHEP2015}. 

The approximate permutation symmetry given in Eq.\ (\ref{Eq:SU5uTeV}) is valid in any flavour basis, and is at the centre of our interest. In the following, we discuss ways to test this $SU(5)$ relation in the case that squarks are observed at the Large Hadron Collider. Note that having degrees of freedom beyond the Standard Model is mandatory in order to test this relation. In particular, if one sends quark masses to infinity, the relation $y_u^t\approx y_u $ alone cannot be tested using SM flavour observables.

% =================================================
\section{An effective theory for the squark sector}

Any strategy that can be set up to test the $SU(5)$ relation Eq.~(\ref{Eq:SU5uTeV}) necessarily relies on a comparison involving at least two up-type squarks. Some of the squarks can be light enough to be produced at the LHC, others may be too heavy such that they appear only virtually in intermediate processes. While the treatment of the eigenvalues of rotation matrices is rather technical, depending on the exact pattern of the up-squark mass matrix, two expansions can be used to in order to simplify the problem: the mass insertion approximation (MIA) or the effective field theory expansion (EFT). The feasability of the $SU(5)$ tests depends crucially on the amount of available data, whether they involve real or virtual squarks.

In many classes of physical models, the squark masses exhibit some hierarchy. The physics of the ``light'' squarks, i.e.\ those accessible at the LHC, can then be captured in an effective Lagrangian, where ``heavy'' squarks are integrated out. The higher-dimensional operators present in this tree-level Lagrangian serve as a basis for the $SU(5)$ tests based on both virtual and real squarks. 

When all up-type squarks are heavier than the typical LHC scale, they can only appear off-shell. Potential tests of the $SU(5)$ hypothesis are thus based on virtual squarks only. All squarks can be integrated out and the effective Lagrangian contains the SM plus possibly other light SUSY particles. If the whole SUSY spectrum is heavy, operators of interest are stemming from one-loop diagrams involving at least one Higgs-squark-squark vertex together with fermions on the outgoing legs in order to access information on the flavour structure. We propose $SU(5)$ tests for this scenario in Refs.\ \cite{PLB2014, JHEP2015}. Moreover, tests based on ultraperiphal searches are proposed in Ref.\ \cite{JHEP2015}.

% =================================================
\section{Application to Natural Supersymmetry}

For the sake of a concrete example, let us consider the case of Natural Supersymmetry. The mass spectrum of this scenario features a first and second generation of squarks that are considerably heavier than those of the third generation. Here, the effective theory will contain two mostly stop-like squarks. Their mixing is not constrained and can potentially be large, which is often needed to satisfy the Higgs-mass constraint. The effective operators that appear when integrating out the heavier squarks can potentially induce flavour-changing stop decays. We assume that both stops are produced at the LHC. The total production cross-section of stop pairs are, at next-to-leading order, $\sigma_{\tilde{t}\tilde{t}} \sim 90$, 8.5, 0.7, and 0.08 fb for stop masses $m_{\tilde{t}} \sim 700$, 1000, 1400, and 1800 GeV, respectively \cite{SUSYXSec}.

In the case where $m_{\tilde{t}_{1,2}} > m_{\tilde{W}} > m_{\tilde{B}}$, the stops may decay either into the lightest ($\tilde{\chi}^0_1 \approx \tilde{B}$) or the second-lightest ($\tilde{\chi}^0_2 \approx \tilde{W}$) neutralino. We are interested in the flavour-changing decays
\begin{equation}
	\tilde{t} ~\to~ \tilde{W} \, u/c ~\to~ \tilde{B} \, Z/h \, u/c \qquad
	\mathrm{and} \qquad
	\tilde{t} ~\to~ \tilde{B} \, u/c \,.
\end{equation}
We assume that the stop masses are unknown and that only the event rates of the above processes, respectively denoted by $N_Y$ and $N_L$, are experimentally accessible. Moreover, we assume that a certain fraction, denoted by $N_Y^c$ and $N_L^c$, can be charm-tagged. We denote $N_L^{\not{c}}$ and $N_Y^{\not{c}}$ the number of events without charm-jets, including those where the charm-tagging fails. Assuming the same charm-tagging efficiency, the relation constituting the $SU(5)$ test is
\begin{equation}
	\frac{N_L^c}{N_Y^c} ~=~ \frac{N_L^{\not{c}}}{N_Y^{\not{c}}}\,.
\end{equation}
The expected precision associated to this test has been evaluated using $p$-value frequentist statistics. We find that testing the relation with 50\% (10\%) precision at $3\sigma$ confidence level requires $N_Y \sim N_L \sim 110$ (2700) events. For comparison, assuming flavour-violating branching ratios of 5\% and an integrated luminosity of 300 fb$^{-1}$, we expect 1340 (11) events for stop masses of 700 (1400) GeV.

For the mass ordering $m_{\tilde{W}} > m_{\tilde{t}_{1,2}} > m_{\tilde{B}}$, the stops can only decay into the bino, according to $\tilde{t} \to t_{R,L} \tilde{B}$. Performing top polarimetry on the decaying stop pair potentially gives acces to the stop mixing angle. The same kind of procedure also provides a $SU(5)$ test. Let us assume that the spin of the tops is analyzed through distributions of the form $(1+\kappa_z)$ with $z \in [-1; 1]$. The decays of the stops $\tilde{t}_a$ and $\tilde{t}_b$ ($a,b=1,2$) leading to the event rates $N_a$ and $N_b$, respectively, are then splitted over the domains $D_- = [-1; 0]$ and $D_+ = [0; 1]$ such that
\begin{equation}
	N_a ~=~ N_a^+ + N_a^- \qquad \mathrm{and} \qquad N_b ~=~ N_b^+ + N_b^- \,.
\end{equation}
These event rates again satisfy a certain (non-trivial) relation if the $SU(5)$ hypothesis is verified. 

Assuming a spin analyzer of efficience $\kappa = 0.5$ and $N_a = 20$ events, we find that $N_b \gtrsim 137$ events are needed to probe the mentioned relation at $3\sigma$ confidence level. Testing the relation with 50\% (20\%) precision requires $N_b \approx 589$ ($7560$) events. For comparison, given an integrated luminosity of $300~\mathrm{fb}^{-1}$, we expect about 26700 (213) events for stop masses of about 700 (1400) GeV.

For a more detailed discussion of this $SU(5)$ test as well as applications to other mass hierarchies see Refs.\ \cite{PLB2014, JHEP2015}.

% =================================================
\section{Conclusion and perspectives}

In summary, we have analyzed how an accidental permutation symmetry present in supersymmetric $SU(5)$-theories may allow to test the $SU(5)$ hypothesis in the case where squarks are observed at the Large Hadron Collider. 

As a first example, we have presented a test based on an effective theory for the squark sector. More tests based on this effective theory and the mass insertion approximation have been published recently. All details and studies for other cases, in particular two-generation and heavy Supersymmetry, can be found in Refs.\ \cite{PLB2014, JHEP2015}.

For a more generic supersymmetric spectrum, other methods turn out to be more efficient. In particular, Bayesian statistics provides a convenient framework  to set up global tests of the approximate permutation symmetries. A first study focussed on LHC observables has been presented in Ref.\ \cite{PhDStoll}, while a more refined analysis is subject to current work.

% =================================================

\end{document}